\documentclass[aps,prl,twocolumn,superscriptaddress,floatfix,showpacs]{revtex4-1}

\usepackage{graphicx}
\usepackage{amsmath}
\usepackage{booktabs}
\DeclareMathOperator{\sech}{sech}

\begin{document}

\title{Dispersive stiffness of Dzyaloshinskii domain walls}

\author{J.P. Pellegren}

\author{D. Lau}

\affiliation{Department of Materials Science \& Engineering, Carnegie Mellon University, Pittsburgh, Pennsylvania 15213, USA}

\author{V. Sokalski}
\email[]{vsokalsk@andrew.cmu.edu}
\affiliation{Department of Materials Science \& Engineering, Carnegie Mellon University, Pittsburgh, Pennsylvania 15213, USA}

\date{\today}

\begin{abstract}

It is well documented that subjecting perpendicular magnetic films which exhibit the interfacial Dzyaloshinskii-Moriya interaction (DMI) to an in-plane magnetic field results in a domain wall (DW) energy, $\sigma$, that is highly anisotropic with respect to the orientation of the DW in the film plane, $\Theta$. We demonstrate that this anisotropy has a profound impact on the elastic response of the DW as characterized by the surface stiffness, $\tilde{\sigma}(\Theta) = \sigma(\Theta) + \sigma''(\Theta)$, and evaluate its dependence on the length scale of deformation. The influence of stiffness on DW mobility in the creep regime is assessed, with analytic and numerical calculations showing trends in $\tilde{\sigma}$ that better represent experimental measurements of domain wall velocity in magnetic thin films compared to $\sigma$ alone. Our treatment provides experimental support for theoretical models of the mobility of anisotropic elastic manifolds and makes progress toward a more complete understanding of magnetic domain wall creep.

\end{abstract}

\maketitle

\begin{figure*}
\includegraphics[width = 7in]{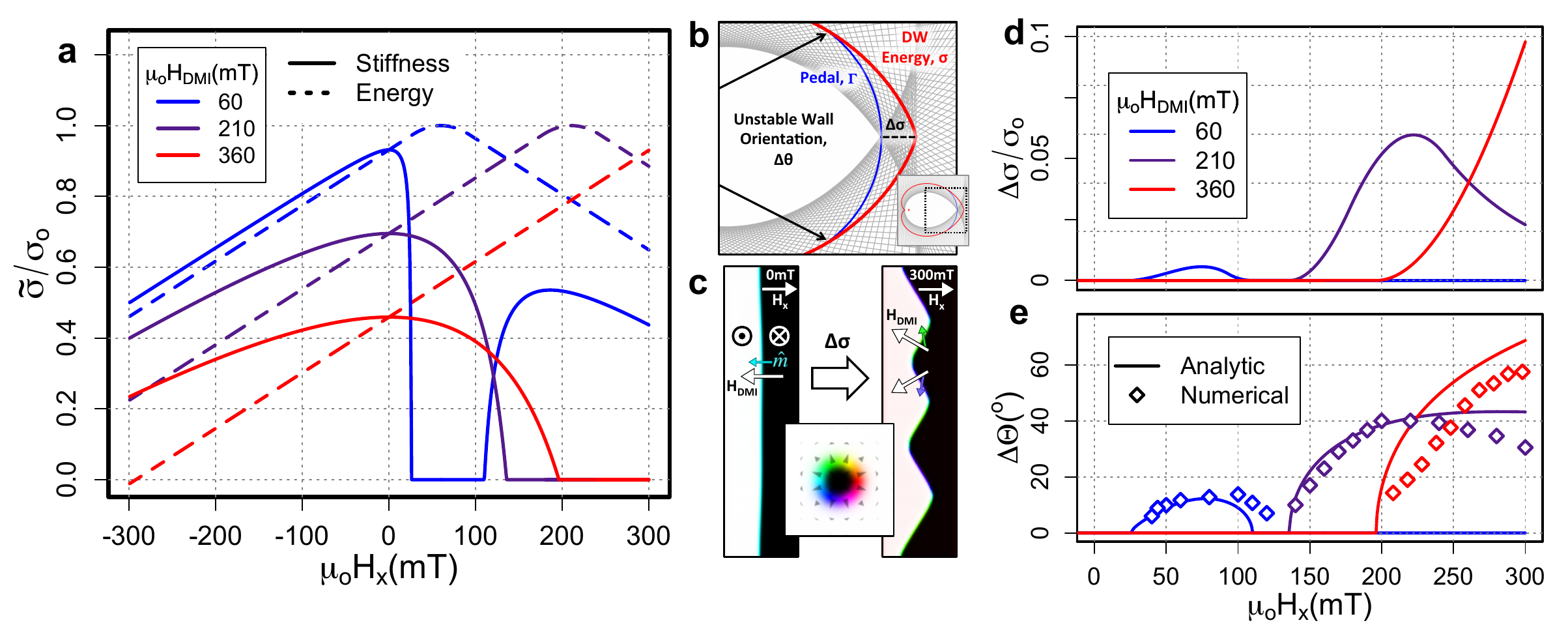}
\caption{\label{InterfaceThermodynamics}a) $\sigma^{eq}$ and long wavelength $\tilde{\sigma}$ vs $\mu_oH_x$ for varying $\mu_oH_{DMI}$ with $\mu_oH_k=$1T, $M_s=600 kA/m$, and $t_f=1.8 nm$. b) Wulff construction (grey), $\sigma^{eq}(\Theta)$ (red), and $\Gamma(\Theta)$ (blue) for a Dzyaloshinski DW with $\mu_oH_{x}=300mT$ and $\mu_oH_{DMI}=360mT$.  The red point of the inset indicates the origin. c) Example calculation of wall faceting in Mumax$^3$ for the same conditions as (b).  d) Calculated driving force for faceting, $\Delta\sigma$. e) Calculated facet orientation, $\Delta\Theta$, with superimposed numerical results.}

\end{figure*}

Topologically protected magnetic features such as skyrmions and chiral domain walls (DWs) have emerged as promising candidates for future spintronic devices due to the unprecedented efficiency by which they can be manipulated with electric current.\cite{Bromberg2014,Emori2013,Ryu2014,Heide2008,Heinze2011,Woo2016} The recently discovered interfacial Dzyaloshinskii-Moriya Interaction (DMI) is critical to the stabilization of these features and greatly influences the energetic symmetry of a DW necessitating that existing models for their behavior be refined.  Because DMI energy scales with the cross-product of neighboring spins, $E=-\textbf{D}\cdot (\textbf{S}_1\times \textbf{S}_2)$, it can only have an effect in systems with structural inversion asymmetry (SIA) typically achieved by sandwiching an ultrathin film with dissimilar materials.\cite{Jue2016,Moriya1960,Thiaville2012} In perpendicular thin films with SIA, the interfacial DMI, $D_{int}$, acts on the internal magnetization of a DW with width, $\lambda$, as described by an effective field, $\mu_oH_{DMI} = D_{int}/(M_s\lambda)$, stabilizing the N\'{e}el configuration with preferred chirality relative to the achiral Bloch configuration, which would otherwise be the ground state.\cite{Chen2013,Hrabec2014,Je2013} In the presence of in-plane magnetic fields, such chiral configurations result in asymmetric growth of perpendicular magnetic bubble domains and a minimum in velocity was widely found to occur at a critical in-plane field suggested to correspond to $H_{DMI}$.\cite{Hrabec2014,Je2013,Kabanov2010,Lavrijsen2015} These arguments were based on a proposed inverse correlation between velocity and DW energy, $\sigma$, due to its appearance in the exponent of the creep law.  We note here that the creep law being used is rooted in the 1D elastic band model where an energy scale, $\varepsilon$, describes the potential associated with bending deformation of the interface. For the case of weak collective pinning, scaling analysis gives

\begin{align}\begin{split} v_{creep} &= v_o e^{\alpha H_z^{-1/4}}\\ \alpha &\propto \varepsilon^{1/4}\end{split}\end{align}

This elastic energy scale is routinely assumed to be given simply by $\varepsilon=\sigma$ for a magnetic domain wall\cite{Lemerle1998,Kim2009},  which is strictly true only for the isotropic case where $\sigma$ is not a function of the wall orientation. The combination of an in-plane field, which alters the wall energy depending on its internal magnetization, and DMI, which couples this magnetization with orientation, produces an anisotropic wall energy. In this case, $\varepsilon=\tilde{\sigma}$ where $\tilde{\sigma}(\Theta) = \sigma(\Theta) + \sigma''(\Theta)$ is the surface stiffness which depends not only on the energy of the local orientation but also on the energies of orientations in close proximity to $\Theta$. Surface stiffness has previously found broad utility in describing the mobility of solid/liquid interfaces\cite{Nozieres1989} and was first employed in creep theory to describe the movement of flux lines through pinning sites in anisotropic superconductors.\cite{Sudbo1991,Blatter1994}

In this letter, we analytically calculate the stiffness of Dzyaloshinskii DWs with anisotropic surface energy, identifying a driving force to spontaneously form facets which we interpret using classical interface thermodynamics and confirm numerically. The impact of symmetric exchange along the domain wall on stiffness is determined over a range of perturbation length scales, allowing for comparison to domain growth experiments. We demonstrate that the creep law predicts the observed asymmetric trends of field-driven wall velocity vs. in-plane field when the elastic properties are taken into account using an additional parameter associated with the length scale of the wall deformation.

Thin films of Pt(2.5)/[Co(0.2)/Ni(0.6)]$_2$/Co(0.2)/ Ta(0.5)/TaN(6) with units in nm were prepared by DC magnetron sputtering on oxidized silicon with working pressure fixed at 2.5mTorr Ar.  Films were determined to have FCC(111) fibre-texture by x-ray diffraction (XRD) with continuous interfaces confirmed by transmission electron microscopy (TEM) as shown in prior work.\cite{Jaris2017}  $K_{eff}$ was determined from the in-plane saturation field, $\mu_oH_k$ (= 1T), by alternating gradient field magnetometry (AGFM) and $M_s = 600kA/m$ determined from vibrating sample magnetometry (VSM).  Studies on domain growth and morphology were performed using a wide-field white light Kerr microscope with an in-plane electromagnet producing static fields up to 300mT and a perpendicular coil producing 1ms pulses up to 20mT.  Bubble domains of 20$\mu m$ diameter were nucleation at points damaged by a Ga$^+$ ion beam as described in \cite{Lau2016}.  Walls were positioned in the field of view by nucleation at the edge of the sample followed by a series of perpendicular field pulses.  Numerical calculations were performed on a mesh of 1 x 1 x 2nm cells using the micromagnetic energy minimization algorithm in Mumax$^3$ version 3.8.\cite{Vansteenkiste2014}. All calculations assumed an exchange stiffness, A = $1\times10^{-11} J/m$ 

In the following analyses, we approximate Dzyaloshinskii DW energy as a function of orientation and internal magnetization as follows:
\begin{multline}\label{eq:sig}
 \sigma(\Theta,\phi)=\sigma _{o}-\pi D_{int}\cos(\phi-\Theta)-\pi\lambda \mu_oH_xM_s\cos(\phi)\\
 +\frac{\ln(2)}{\pi}t_f\mu_oM_s^2\cos^2(\phi-\Theta)
\end{multline}

Where, $\Theta$ and $\phi$ represent the azimuthal angles of the DW normal and internal magnetization $\textbf{m}$, respectively. The direction of the applied in-plane field, $H_x$, defines the x-axis while the effective field due to DMI is always oriented along the DW normal. The constants $\sigma_o=4\sqrt{AK_{eff}}$ and $\lambda=\sqrt{A/K_{eff}}$ are the Bloch wall energy and width.  The fourth term is the DW anisotropy energy rooted in the magnetostatic favorability of Bloch walls over N\'{e}el walls.\cite{Thiaville1995} The validity of equation \ref{eq:sig} is confirmed through comparison with numerical calculations of DW configuration to follow.

Minimizing DW energy with respect to internal magnetization results in the equilibrium values for a rigid wall as a function of orientation, $\sigma^{eq}(\Theta)$ and $\phi^{eq}(\Theta)$, which have been used in the calculation of DW tilting angles in nanowires and equilibrium droplet shapes via the Wulff construction.\cite{Wulff1901,Herring1951} A stiffness value can be calculated from this energy simply by

\begin{equation}\label{eq:lwstiff}{\tilde\sigma(\Theta,L\to\infty)=\sigma^{eq}(\Theta)+\frac{\partial^2\sigma^{eq}}{\partial\Theta^2}(\Theta)}\end{equation}

As noted above and discussed later in the text, this expression is valid only for long wavelength distortion and so provides insight into the long range stability of the planar domain wall. In Figure \ref{InterfaceThermodynamics}a, $\sigma^{eq}$ and the long wavelength limit of $\tilde{\sigma}$ vs. $\mu_oH_x$ are compared for Dzyaloshinskii DWs with varying $H_{DMI}$, where positive $H_x$ implies it is anti-parallel to $H_{DMI}$ (henceforth referred to as the anti-parallel case).  While $\sigma^{eq}$ is symmetric about a maximum at $H_{DMI}$, $\tilde{\sigma}$ is highly asymmetric about a maximum centered at $H_x=0$. Although $\sigma^{eq}$ and $\tilde{\sigma}$ are qualitatively similar when the applied field is parallel to $H_{DMI}$, we note the striking result that as $\sigma$ increases for anti-parallel $H_x$, $\tilde{\sigma}$ drops rapidly at a field which is dependent on $H_{DMI}$. The calculation of $\tilde{\sigma}$ is complicated by the occurrence of negative values around $H_{DMI}$, as they suggest that a non-planar, faceted wall configuration is favored. The thermodynamic properties of faceted configurations have been explored in the study of crystal growth and are described geometrically by the pedal, $\Gamma(\Theta)$, of the equilibrium profile as determined via the Wulff construction on the polar energy plot, $\sigma(\Theta)$ (Figure \ref{InterfaceThermodynamics}b). The pedal gives both the driving force for faceting, $\Delta\sigma$, and the facet angles, $\Delta\Theta$, as indicated.\cite{Burton1951}\cite{Mullins1962} The faceting angle becomes non-zero where $\tilde{\sigma}$ is negative, and optimum orientations calculated from equation \ref{eq:sig} show good agreement with numerical energy minimization results (Figure \ref{InterfaceThermodynamics}e).\cite{Vansteenkiste2014} Calculating stiffness from $\Gamma(\Theta)$, we find that the linear elastic response vanishes for cases where faceting is favored.

In real materials, a pinning potential can deform the domain wall at small length scales, in which case Heisenberg exchange along the wall will prevent the internal magnetization from assuming $\phi^{eq}(\Theta)$ everywhere and invalidate equation \ref{eq:lwstiff}. In order to express the stiffness in this case, the magnetization profile and energy for an infinitesimally curved domain wall can be calculated semi-analytically by considering perturbations of the 1-D expression in equation \ref{eq:sig}. We consider a narrow domain wall, neglecting non local terms in the demagnetizing energy. Augmenting the 1-D energy with an additional term for the exchange along the DW\cite{Supplemental1}, the combined energy functional is

\begin{equation}\label{eq:engfun}{\mathcal{E}=\int\sigma(\Theta,\phi)+2A\lambda\left(\frac{\partial \phi}{\partial s}\right)^2ds}\end{equation}

We expand the 1-D energy to second order about the orientation, $\Theta_o$, and equilibrium magnetization, $\phi_o$, for a straight DW segment
\begin{multline}\label{eq:sig_exp}\sigma(\Theta,\phi)=\sigma+(\Theta-\Theta_o)\sigma_\Theta+(\phi-\phi_o)(\Theta-\Theta_o)\sigma_{\Theta\phi}\\
+\frac{1}{2}(\Theta-\Theta_o)^2\sigma_{\Theta\Theta}+\frac{1}{2}(\phi-\phi_o)^2\sigma_{\phi\phi}\end{multline}

In the right hand side of the above equation, $\sigma$ and its partial derivates, indicated by subscripts, are all evaluated at $(\Theta_o,\phi_o)$. A stationary $\phi$ profile will satisfy the Euler-Lagrange equation 
\begin{equation}\label{eq:E-L}{(\Theta-\Theta_o)\sigma_{\Theta\phi}+(\phi-\phi_o)\sigma_{\phi\phi}-4A\lambda\frac{\partial^2\phi}{\partial s^2}=0}\end{equation}

A segment that is deformed into a circular arc of radius $R$ has an orientation profile along the wall given by
\begin{equation}{\Theta(s)=\Theta_o - \frac{s}{R}}\end{equation}

Which can be used to solve equation \ref{eq:E-L} for the magnetization profile, giving

\begin{align}\label{eq:phi}\begin{split}\phi(s)=\phi_o + \frac{s}{R}\frac{\sigma_{\Theta\phi}}{\sigma_{\phi\phi}} + &C_1\sinh\left(\frac{s}{\Lambda}\right)+C_2\cosh\left(\frac{s}{\Lambda}\right)
\\ \Lambda &= \lambda\sqrt\frac{\sigma_o}{\sigma_{\phi\phi}}\end{split}\end{align}

Here $\Lambda$ is the length scale for exchange along the domain wall, the vertical Bloch line width. For a domain wall segment with fixed endpoints a length $L$ apart, the bounds of integration for large $R$ are
\begin{equation}\label{eq:ep}{s_{ep}=\pm R\arcsin\left(\frac{L}{2R}\right)\approx\pm\frac{L}{2}\left(1+\frac{1}{6}\left(\frac{L}{2R}\right)^2\right)}\end{equation}

Combining the expressions for $\sigma$, $\Theta$, $\phi$, and $s_{ep}$, we evaluate $\mathcal{E}$ of the curved segment and will focus on the case where we do not fix the magnetization of the endpoints. Minimizing energy with respect to $C_1$ and $C_2$ we have
\begin{align}\begin{split}C_1 &= \frac{\sigma_{\Theta\phi}}{\sigma_{\phi\phi}}\frac{\Lambda}{2R}\sech{\frac{L}{2\Lambda}}\\
C_2 &= 0\end{split}\end{align} 

The ground state energy of a curved domain wall can now be directly determined, from which we can extract the elastic response through the relation 
\begin{equation}{\mathcal{E}(R)\approx L\left(\sigma + \frac{1}{6}\left(\frac{L}{2R}\right)^2\tilde\sigma\right)}\end{equation}

The result is a dispersive stiffness given by
\begin{align}\begin{split}\tilde\sigma(\Theta,L)&=\sigma+\sigma_{\Theta\Theta}-\frac{\sigma_{\Theta\phi}^2}{\sigma_{\phi\phi}}\zeta\left(\frac{L}{2\Lambda}\right) \\ \zeta(\ell) &= 1-\frac{3}{\ell^3}\left(\ell-\tanh(\ell)\right)\end{split}\end{align}

\begin{figure}
\includegraphics[width = 3.4in]{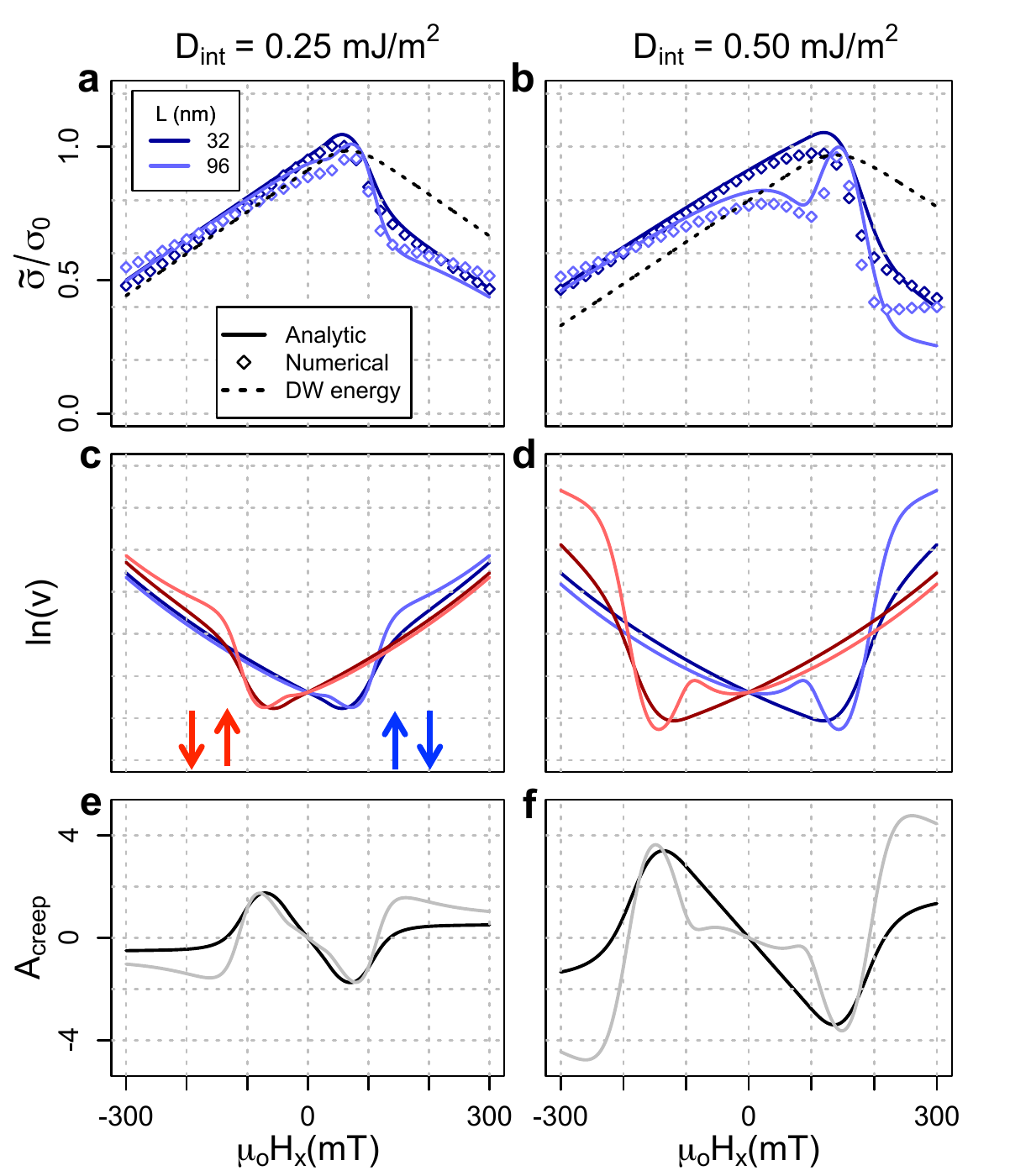}
\caption{\label{StiffvsL} a-b) Normalized $\tilde{\sigma}$ vs $\mu_oH_x$ for a) $D = 0.25 mJ/m^2$ and b) $D = 0.5mJ/m^2$ from analytic and 2D numerical calculations.  Dashed lines indicate DW energy. c-d) Corresponding $\uparrow\downarrow$ and $\downarrow\uparrow$ DW velocity behavior calculated from a,b. e-f) Anti-symmetric component of the velocity, $A_{creep}=\ln(v_{\uparrow\downarrow}/v_{\downarrow\uparrow})$ calculated from c,d.} 
\end{figure} 

As $L\to\infty$, $\zeta\to1$ and we recover an expression for $\tilde{\sigma}$ that is independent of symmetric exchange along the DW.  This expression is the generalized stiffness for a surface with an orientational order parameter in local equilibrium that was first identified by Fournier to describe "soft" materials.\cite{Fournier1995}  Conversely, as $L\to0$, $\zeta\to0$ and the stiffness corresponds to the domain wall bending while maintaining a constant internal magnetization direction. The third term in the expression for stiffness therefore corresponds to the energy decrease due to the DW moments relaxing to the ground state of the curved wall.   

Stiffness as a function of $\mu_oH_x$ is plotted in Figure \ref{StiffvsL} for different L and $D_{int}$ with $\mu_oH_k=1T$ and $\Theta_o$ fixed at $8^\circ$ to account for roughness as justified later. 2D numerical calculations have been superimposed and show good agreement with the analytic solution as described in the supplemental information. To better compare the trends in $\tilde\sigma$ with experiment, we have calculated a DW velocity, $v$, and its anti-symmetric component, $A_{creep}$, as defined in \cite{Jue2016} with the exponential factor scaling by $(\tilde\sigma/\tilde\sigma_{H_x=0})^{1/4}$. We see from either $\tilde\sigma$ or velocity that at low length scales the effect of DMI is to both shift the curve horizontally, as described by Je \textit{et al.}\cite{Je2013} for $\sigma$, and induce an asymmetric vertical shift which is superficially similar to the chiral damping proposed by Ju\'e \textit{et al.}\cite{Jue2016}. Unlike chiral damping, these two effects offset each other at high fields so the stiffness converges for $\uparrow\downarrow$ and $\downarrow\uparrow$ walls. As L increases, sharp drops in stiffness develop at fields where the wall transitions from fully N\'{e}el to having some Bloch component. The most striking consequence is that the anti-parallel case can have multiple local extrema as well as a significant window where it is expected to have a greater velocity than its parallel counterpart before the two cases converge at much larger fields. 

\begin{figure}[ht]
\includegraphics[width = 3.4in]{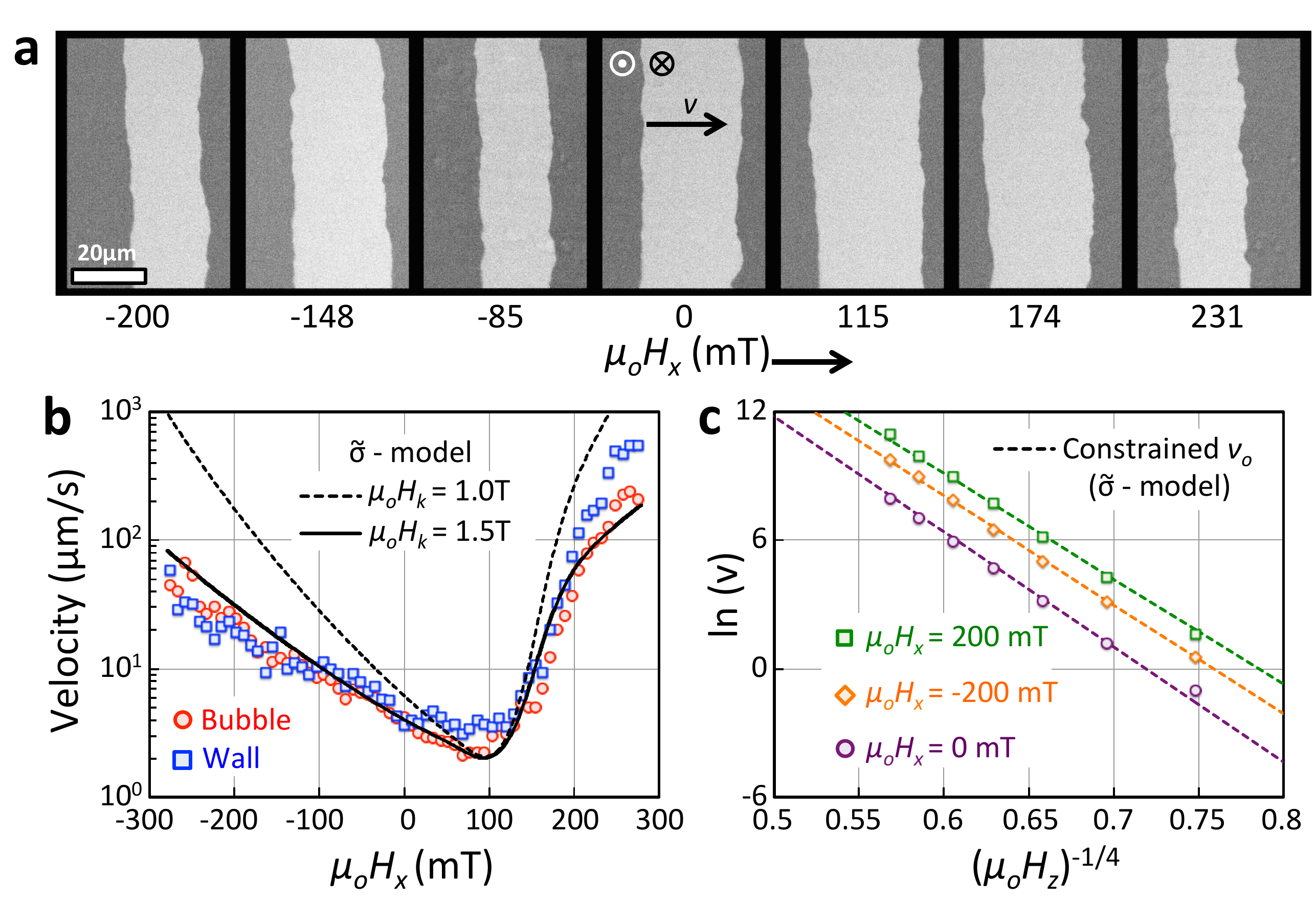}
\caption{\label{Kerr_Images} a) Subtractive Kerr images of Co/Ni where bounds of the light grey region represent the displacement of an $\uparrow\downarrow$ DW. b) Experimental velocity vs. $\mu_oH_x$ for $\mu_oH_z=4.25mT$ with fit from the dispersive stiffness model. The fitting parameters use $v_o$ and $\alpha$ taken from experimental results at $H_x=0$. c) DW velocity vs $(\mu_oH_z)^{-1/4}$ for a series of $\mu_oH_x$. Dashed lines correspond to creep parameters used in our stiffness model where $v_o$ is fixed.}
\end{figure}

We now turn our attention to experimental studies on films with SIA supporting the argument for stiffness as the governing factor in Dzyaloshinskii DW mobility based on the previous calculations. In recent work, the anti-parallel case was found to grow with greater velocity than its parallel counterpart for large $H_x$\cite{Lau2016,Wells2017}, which is contradictory to any explanation based on a velocity trend that is inversely related to $\sigma$. In the case of Co/Ni, this was attributed to an energetic driving force to form the equilibrium shape.\cite{Lau2016} However, we have found that the velocity trends are largely independent of the bubble size in this system and hold even for planar DWs, suggesting that the mechanism is more fundamental to the domain wall mobility. In Figure \ref{Kerr_Images}, we observe a distinct asymmetry in the Co/Ni system characterized by a rapid increase in velocity for large positive $H_x$ and a more gradual monotonic increase for negative $H_x$ noting that there is little difference between the trends for bubble domains or planar DWs.  A series of Kerr images are included to highlight the evolution of the wall morphology with increasing $H_x$. Although there are likely to be changes to the morphology on a length scale not resolvable by Kerr microscopy, it does appear that the wall profile becomes more irregular for increasing $H_x$ that could be due to a reduction of stiffness. 

The experimental data is best fit using the stiffness model with $\mu_oH_{DMI}=106mT$, $L = 47nm$, $\mu_oH_{k}=1.5T$, and $\Theta_o = 8^\circ$.  Among these fitting parameters, each has a markedly different effect on the shape of the curve making it infeasible to maintain a fit to the experimental data by simultaneously varying multiple values.  On this front, our model not only allows us to determine the magnitude of $D_{int}$, but also provides an estimate of the pinning length scale; something that is notoriously difficult to extract experimentally.  The non-zero value of $\Theta_o$ accounts for inherent roughness of the DW and is consistent with the wall profile as seen in Figure \ref{Kerr_Images}a.  Although the fit suggests an anisotropy field larger than we observe experimentally, this could be explained by a field dependence of the pinning potential or domain wall width, which can both affect the creep law energy scale in addition to the elastic properties.\cite{Blatter1994}

We also note that the value of $H_{DMI}$ predicted here is larger than the field at which the minimum in velocity is observed. Although a relatively small deviation here, it could be significant in systems where L is large leading to a local maximum in velocity with an initial minimum shifted closer to the origin as in, for example, Figure \ref{StiffvsL}d with L = 96nm. Indeed, a recent study on Hf/FeCoB/MgO thin films identified a local maximum in velocity for small $H_x$, which matches this trend and is not predicted from other theoretical treatments.\cite{Soucaille2016} In low coercivity films such as FeCoB/MgO, it is reasonable to expect the pinning sites to be more sparse leading to larger values of L.  

In summary, we have demonstrated that the property governing Dzyaloshinskii DW mobility is $\tilde{\sigma}$ as predicted by the elastic band model of creep\cite{Sudbo1991,Blatter1994,Lemerle1998,Kim2009} rather than the DW energy, $\sigma$, as often assumed. Replacing $\sigma$ with $\tilde{\sigma}$ in the exponent of the creep law is inconsequential to the study of DW dynamics absent a symmetry breaking in-plane magnetic field, but critical when the effects of such a field are combined with interfacial DMI. This model explains multiple features of the velocity curves including a reversal in growth symmetry and a local maximum in velocity that are not predicted from past treatments.  By fitting experimental data, it is possible to not only extract $D_{int}$, but also the length scale of the DW deformation. We note that while the elastic energy has typically been applied within the exponential of the creep law, it is possible that the stiffness would affect the frequency of thermal vibrations of the DW, which in turn could impact the attempt frequency and therefore the pre-exponential factor. Although our analysis is distinctly different from previous descriptions of Dzyaloshinskii DW creep, the modification is not to the creep law itself, but to built-in assumptions about wall elasticity.  This treatment actually reaffirms the broad validity of describing magnetic domain walls as 1D elastic bands in thin films.   

\begin{acknowledgments}

The authors are grateful for support of this research from the Samsung Global MRAM Innovation Project.  This work was funded (in part) by the Dowd Fellowship from the College of Engineering at Carnegie Mellon University. The authors would like to thank Philip and Marsha Dowd for their financial support and encouragement.  J.P.P. acknowledges support from The John and Claire Bertucci Fellowship through the College of Engineering at Carnegie Mellon University.

\end{acknowledgments}

\bibliography{Nature_Bibliography}

\end{document}